\documentclass[aps, prd, superscriptaddress, nofootinbib, preprintnumbers, showpacs]{revtex4}
\usepackage{amsmath}
\usepackage{graphicx}

\newcommand{\lesssim}{\mathrel{\mathpalette\vereq<}}

\begin{document}
\preprint{MISC-2011-08}
\title{%
  \hfill{\normalsize\vbox{%
  }}\\
  \vspace{-0.5cm}
  {\bf  Anomalous $\omega$-$Z$-$\gamma$ Vertex from Hidden Local Symmetry} } 
\author{Masayasu Harada}\thanks{
      {\tt harada@hken.phys.nagoya-u.ac.jp}}
       \affiliation{ Department of Physics, Nagoya University,
                    Nagoya, 464-8602, Japan.}
\author{Shinya Matsuzaki}\thanks{
      {\tt synya@cc.kyoto-su.ac.jp} }
      \affiliation{ Maskawa Institute for Science and Culture, Kyoto Sangyo University, Motoyama, Kamigamo, Kita-Ku, Kyoto 603-8555, Japan.}
\author{{Koichi Yamawaki}} \thanks{
      {\tt yamawaki@kmi.nagoya-u.ac.jp}}
      \affiliation{ Kobayashi-Maskawa Institute for the Origin of Particles and 
the Universe (KMI) \\ 
 Nagoya University, Nagoya 464-8602, Japan.}
\date{\today}

\begin{abstract} 
We formulate the general form of 
$\omega$-$Z$-$\gamma$ vertex in the framework based 
on the hidden local symmetry (HLS), which arises
 from the gauge invariant terms for intrinsic parity-odd (IP-odd) part of the effective action.
Those terms are given as the  homogeneous part of the general solution (having free parameters) to 
the Wess-Zumino (WZ) anomaly equation and hence are not determined by the anomaly, 
in sharp contrast to the Harvey-Hill-Hill (HHH) action where  
the relevant vertex is claimed to be uniquely determined by the anomaly.
We show that, even in the framework that HHH was based on, 
 the $\omega$-$Z$-$\gamma$ vertex  
is actually not determined by the anomaly but by the homogeneous (anomaly-free) part of the general solution to the WZ anomaly equation 
having free parameters  
in the same way as in the HLS formulation: The HHH action is just 
a particular choice of the free parameters in the general solution. 
We further show that 
the $\omega$-$Z$-$\gamma$ vertex related to 
the neutrino ($\nu$) - nucleon ($N$) scattering cross section 
$\sigma(\nu N \rightarrow \nu N (N') \gamma)$ 
is determined not by the anomaly
but by the anomaly-free part of the general solution having free parameters.  
Nevertheless we find that the cross section $\sigma(\nu N \rightarrow \nu N (N') \gamma)$ 
is  related through the Ward-Takahashi identity  to 
$\Gamma(\omega \rightarrow \pi^0 \gamma)$ which has  the same parameter-dependence as that of  $\sigma(\nu N \rightarrow \nu N (N') \gamma)$ 
and hence the ratio  $\sigma(\nu N \rightarrow \nu N (N') \gamma)/\Gamma(\omega \rightarrow \pi^0 \gamma)$ is fixed  independently  of these
free parameters. 
Other set of the free parameters of the general solution can be fixed  
to make the best fit of the $\omega \rightarrow \pi^0 l^+ l^-$ process, 
which substantially differs from the HHH action.  This gives a  prediction of
the cross section $\sigma(\nu N \to \nu N (N') \gamma^*(l^+ l^-))$  
to be tested at $\nu$-$N$ collision experiments in the future. 
\end{abstract}

\maketitle

\section{Introduction}

The chiral Lagrangian describes  
the low-energy properties of QCD governed by the low-lying hadron
spectrum including the pseudo Nambu-Goldstone  
bosons (NGBs) associated with the spontaneous chiral symmetry
breaking.  
In this framework Ward-Takahashi identities of QCD play an essential 
role to determine forms of interactions between the NGB and 
external gauge fields arising from gauging the chiral symmetry. 
 Among those symmetry identities, the anomalous Ward-Takahashi identity 
is of great importance,  which is tied with non-Abelian anomaly arising from 
the underlying quark currents in QCD. 
It is reproduced in the chiral Lagrangian by (covariantized)
Wess-Zumino-Witten (WZW) term 
$\Gamma_{\rm WZW}$~\cite{Wess:1971yu,Witten:1983tw}, 
which satisfies the non-Abelian anomaly equation, called the Wess-Zumino (WZ) anomaly equation, 
\begin{eqnarray} 
 \delta \Gamma [{\mathcal L},{\mathcal R}, U] =\delta \Gamma_{\rm QCD} &=&  
- \frac{N_c}{24 \pi^2} \int_{M^4} {\rm tr}\left[
\epsilon_L \left\{  (d {\mathcal L})^2 
- \frac{i}{2} d {\mathcal L}^3 \right\}
 - ({\mathcal L} \leftrightarrow {\mathcal R}) 
\right] 
\equiv {\bf A}
\,, \label{WZeq}
\end{eqnarray} 
where $M^4$ denotes four-dimensional Minkowski manifold,  and 
$U$, ${\mathcal L}$ and ${\mathcal R}$ denote the chiral 
field parameterizing the NGBs $(\pi)$ as $U=e^{2i \pi/F_\pi}$ 
and external gauge fields, respectively.  
Hence the 
low-energy interactions among the NGBs and external gauge fields
such as $\pi^0 \rightarrow \gamma\gamma$ and
$\gamma^* \rightarrow \pi^0\pi^+\pi^-$
are completely determined by the anomaly (Low Energy Theorem).
How about inclusion of the vector mesons ($\omega$, $\rho$, etc.)
in the intrinsic parity odd (IP-odd) processes such as
$\omega$-$Z$-$\gamma$, $\omega$-$\pi^0$-$\gamma$, 
$\omega$-$\pi^0$-$\pi^+$-$\pi^-$, etc. without affecting the Low Energy Theorem? This problem was solved long time ago~\cite{Fujiwara:1984mp} 
in the framework of the Hidden Local Symmetry (HLS)~\cite{Bando:1987br,Bando:1984ej,Harada:2003jx}  which incorporates 
 the vector mesons as the (composite) gauge bosons of HLS in the nonlinear sigma model.
The resulting expression~\cite{Fujiwara:1984mp} implicitly  contained the  $\omega$-$Z$-$\gamma$ vertex which was not analyzed so far.

The method was based on the fact that since
the WZ anomaly equation (\ref{WZeq}) is an inhomogeneous 
linear differential equation,   
its general solution is given by the linear combination of a special
solution to Eq.(\ref{WZeq}), $\Gamma_{\rm WZW}$,
and the general solution to the homogeneous part of
Eq.(\ref{WZeq}) ($\delta \Gamma =0$).   
While the special solution 
$\Gamma_{\rm WZW}$  
is completely fixed by the anomaly,
the general  solution to the homogeneous part $\delta \Gamma=0$
cannot be fixed by the anomaly. 
The general solution $\Gamma_{\rm HLS}^{\rm inv}$ to the homogeneous WZ equation must be gauge-invariant (anomaly-free) and
has actually been found~\cite{Fujiwara:1984mp}:
\begin{equation}
\Gamma_{\rm HLS}^{\rm inv}=  \frac{N_c}{16\pi^2} \int_{M^4}  \sum_{i=1}^{4} c_i {\mathcal L}_i\, ,\quad \delta \Gamma_{\rm HLS}^{\rm inv}= 0\,,
\label{invariant}
\end{equation}
where the explicit expression of ${\mathcal L}_i$ will be given later.
Then the  general solution 
$\Gamma_{\rm HLS}^{\rm full}$ to the WZ anomaly equation Eq.(\ref{WZeq})
is given by the sum of the special solution $\Gamma_{\rm WZW}$ and the gauge invariant terms in Eq.(\ref{invariant}), 
$\Gamma_{\rm HLS}^{\rm inv}=\Gamma_{\rm HLS}^{\rm inv}[V, {\mathcal L} , {\mathcal R} , \xi_L^\dag \xi_R]$,  
having odd-property under intrinsic parity (IP-odd): 
\begin{eqnarray}
\Gamma_{\rm HLS}^{\rm full}
[V, {\mathcal L} , {\mathcal R} , \xi_L^\dag \xi_R]
&=&
\Gamma_{\rm WZW}
[{\mathcal L} , {\mathcal R} , U]
 + \Gamma^{\rm inv}_{\rm HLS}
[V, {\mathcal L} , {\mathcal R} , \xi_L^\dag \xi_R]
\ , 
\label{full:HLS}
\end{eqnarray}
where
$U$ was decomposed as 
$U=\xi_L^\dag \xi_R$~\cite{Bando:1987br,Bando:1984ej,Harada:2003jx}
and  the vector mesons $(V)$
as the gauge bosons of the HLS were involved.
This 
gauge invariant part
$\Gamma_{\rm HLS}^{\rm inv}$
fully describes 
the $\omega$-$Z$-$\gamma$ vertex 
as well as the other IP-odd hadronic processes such as
$\omega$-$\pi^0$-$\gamma$ and $\omega$-$\pi^0$-$\pi^+$-$\pi^-$ 
which were intensively studied in 
Refs.~\cite{Fujiwara:1984mp,Bando:1987br,Harada:2003jx}.
It is remarkable that 
these processes 
are
{\it not} determined by the anomaly,
but by 
the 
{\it invariant} part
$\Gamma_{\rm HLS}^{\rm inv}$
which contains free parameters $c_1$-$c_4$.

On the other hand, 
it has recently been advocated 
by Harvey-Hill-Hill (HHH)~\cite{Harvey:2007rd,Harvey:2007ca}  
that the interactions such as $\omega$-$Z$-$\gamma$ are 
uniquely determined by the non-Abelian anomaly in the presence of 
``background fields'' $B_L$ and $B_R$ 
for vector and axialvector mesons  
in addition to the external gauge fields ${\mathcal L}$ and 
${\mathcal R}$.  Based on the resultant action (``HHH action''), 
they further suggested
that the $\omega$-$Z$-$\gamma$ vertex can explain an excess of 
electron-like events in a low-energy region of neutrino-nucleon
collision processes observed 
at the Fermilab Booster Neutrino Experiment
MiniBooNE~\cite{AguilarArevalo:2007it,AguilarArevalo:2008rc,Djurcic:2009ds}. 

In this paper,  we shall first present explicit form of the  $\omega$-$Z$-$\gamma$ vertex arising from the gauge-invariant (anomaly-free)  term in the HLS formalism.
We then clarify the claim of HHH that the  $\omega$-$Z$-$\gamma$ vertex is determined by the anomaly, which is in obvious contradiction with the HLS formalism.

The HHH incorporated 
the ``background fields'' $B_L$ and $B_R$
transforming  
homogeneously under the chiral (gauge) symmetry
into the 
WZW term, $\Gamma_{\rm WZW}
  [ {\mathcal L}  \,,\, {\mathcal R}  \,,\, U]$,
in such a way that ${\mathcal L} $ and ${\mathcal R} $ are simply
replaced as ${\mathcal L} \rightarrow {\mathcal L}  + B_L$ and
${\mathcal R}\rightarrow {\mathcal R}  + B_R$:
\begin{equation}
\Gamma_{\rm HHH}
  \left[{\mathcal L} , {\mathcal R} , B_L,B_R,U \right] 
= \Gamma_{\rm WZW} [{\mathcal L} +B_L, {\mathcal R}  + B_R, U]
\,. 
\end{equation}
They included the counterterm
$\Gamma_c[{\mathcal L} ,{\mathcal R} ,B_L.B_R]$,
what they call 
``generalized Bardeen's counterterm'',
so  that the full action under the presence of
$B_L$ and $B_R$,
\begin{equation}
\Gamma^{\rm full}_{\rm HHH}
  \left[{\mathcal L} , {\mathcal R} , B_L,B_R,U \right] 
\equiv \Gamma_{\rm HHH}
  \left[{\mathcal L} , {\mathcal R} , B_L,B_R,U \right] 
+ \Gamma_c \left[{\mathcal L} ,{\mathcal R} ,B_L,B_R\right]
\ ,
\end{equation}
satisfies the WZ anomaly equation Eq.(\ref{WZeq}): 
$\delta \Gamma_{\rm HHH}^{\rm full} = {\bf A}$. 
They claimed that counter term  $\Gamma_c$ should be uniquely determined by the anomaly  through Eq.(\ref{WZeq}). 
This would imply that 
the difference of them
 \begin{eqnarray}  
\Delta \Gamma_{\rm HHH}[{\mathcal L} , {\mathcal R} , B_L,B_R,U] 
\equiv  
\Gamma_{\rm HHH}^{\rm full} 
  \left[{\mathcal L} , {\mathcal R} , B_L,B_R,U \right] 
-
\Gamma_{\rm WZW}[{\mathcal L} ,{\mathcal R} ,U]
\, 
 \label{HHH:WZW}
\end{eqnarray} 
should also be determined by the anomaly.

Note, however, that 
both 
$\Gamma^{\rm full}_{\rm HHH}
$ and $\Gamma_{\rm WZW} 
(= \Gamma_{\rm HHH}
 \left[B_L=B_R=0\right] )$ in Eq.(\ref{HHH:WZW}) 
satisfy the same anomaly equation (\ref{WZeq}), i.e., 
$\delta \Gamma^{\rm full}_{\rm HHH}=\delta \Gamma_{\rm WZW} = {\bf A}$, 
 and hence $\Delta \Gamma_{\rm HHH}$   
should be invariant (anomaly-free) under the gauge
transformation:
\begin{equation}
\delta \left(
\Delta \Gamma_{\rm HHH} [{\mathcal L} , {\mathcal R} , B_L,B_R,U] 
\right) 
= 0 
\ ,
\label{HHH}
\end{equation}
in contradiction with the HHH claim. Actually, 
there are a lot of solutions which satisfy Eq.(\ref{HHH}).  
We shall call the general solution to Eq.(\ref{HHH})
$\Gamma_{\rm G-HHH}^{\rm inv}$ (G-HHH action), 
\begin{equation} 
 \delta \, \Gamma_{\rm G-HHH}^{\rm inv} = 0 
\,,\label{HHH:inv:eq}
\end{equation}
which is not determined by the anomaly, precisely the same situation as 
$\Gamma_{\rm HLS}^{\rm inv}$ in the HLS formulation.

In order to make the above our argument more explicit, 
we shall present the general solution to Eq.(\ref{HHH:inv:eq}), 
the G-HHH action $\Gamma_{\rm G-HHH}^{\rm inv}$, 
in the framework that HHH was based on. (We call it ``HHH formulation"). 
It turns out that $\Gamma_{\rm G-HHH}^{\rm inv}$ is given by 
a linear combination of fourteen chiral (gauge) invariant IP-odd terms, 
which clarifies that the definite form of the HHH action  
$\Delta \Gamma_{\rm HHH}$ is actually a {\it particular} 
choice of the general solution $\Gamma_{\rm G-HHH}^{\rm inv}$.

To be concrete, we next discuss $\omega$-$Z$-$\gamma$ vertex.
 We formulate the general form of $\omega$-$Z$-$\gamma$ vertex 
arising from the gauge invariant HLS action $\Gamma_{\rm HLS}^{\rm inv}$ 
in Eq.(\ref{invariant})
as well as the G-HHH action $\Gamma_{\rm G-HHH}^{\rm inv}$ having 
free parameters.  In spite of the free parameters,  certain combinations of the physical quantities can be fixed
independently of these free parameters by taking the ratio having the same parameter-dependence.  
We find that the $\omega$-$Z$-$\gamma$ vertex is related 
to the $\omega$-$\pi^0$-$\gamma$ vertex 
through the Ward-Takahashi identity. 
We evaluate contributions from the $\omega$-$Z$-$\gamma$ vertex to a neutrino ($\nu$) - nucleon ($N$) 
cross section $\sigma(\nu N \to \nu N^{(')} \gamma)$ 
using the experimental input for the $\omega \to \pi^0 \gamma$ decay. 
 Furthermore, existence of other free parameters in the general solution enables us to make the best fit 
of the $\omega$-$\pi^0$-$\gamma^*$ process, which is substantially different from the HHH action.
Based on the best fit parameter choice, we give a prediction of 
the  cross section $\sigma(\nu N \to \nu N \gamma^*)$ 
to be tested at $\nu$-$N$ collision experiments in the future.

This paper is organized as follows: 
In Sec.~\ref{wZg-vertex} 
we derive the explicit expression of 
$\omega$-$Z$-$\gamma$ vertex arising from the gauge invariant IP-odd terms in the HLS formulation.  
In Sec.~\ref{comp-HHH}  to make a comparison of the HLS result with the HHH one,  
we show that the $\omega$-$Z$-$\gamma$ vertex is not determined by the anomaly but by the general solution to Eq.(\ref{HHH:inv:eq}), the G-HHH action $\Gamma_{\rm G-HHH}^{\rm inv}$,  which also has free parameters, based on the same framework
that HHH was based on.  We then demonstrate that the HHH action $\Delta \Gamma_{\rm HHH}$ 
is nothing but a particular choice of $\Gamma_{\rm G-HHH}^{\rm inv}$. 
In Sec.~\ref{implications} 
we discuss phenomenological applications associated with 
the $\omega$-$Z$-$\gamma$ vertex: 
$\sigma(\nu N \to \nu N \gamma)$ 
and $\sigma(\nu N \to \nu N \gamma^* (l^-l^+))$.  
Summary is devoted to Sec.~\ref{summary}. 
In Appendix~\ref{relation} we show the explicit relation between the HLS and HHH action 
by integrating out the axialvector mesons of the HHH action. 
In Appendix~\ref{GHLS:inv} 
we will also show the relation between 
the general solution $\Gamma_{\rm G-HHH}^{\rm inv}$ and 
the IP-odd gauge invariant terms~\cite{Kaiser:1990yf} formulated in 
the generalized HLS (GHLS)~\cite{Bando:1984pw,Bando:1987br,Bando:1987ym}.

\section{The $\omega$-$Z$-$\gamma$ vertex in the HLS formalism} 
\label{wZg-vertex}

 We begin by briefly reviewing 
the HLS formalism~\cite{Bando:1987br,Bando:1984ej,Harada:2003jx} 
to introduce  the explicit form of the gauge  invariant IP-odd 
terms in $\Gamma^{\rm inv}_{\rm HLS}$ of Eq.(\ref{invariant})~\cite{Fujiwara:1984mp}.  
 The basic dynamical variables are nonlinear base $\xi_{L,R}(x)$ embedded into 
the chiral field $U(x)$ as $U(x)=\exp (2 i\pi(x)/F_\pi)=\xi_L^\dagger(x) \xi_R(x)$ 
associated with the spontaneously breaking of global chiral symmetry $G_{\rm global}=U(N_f)_L \times U(N_f)_R$ down to 
$H_{\rm global}=U(N_f)_{V=L+R}$, 
where $\pi(x)$ denotes the NGB fields having the decay constant $F_\pi$ and  
$N_f$ the number of massless quark flavors 
(quark mass is disregarded throughout this paper).  
The chiral field $U$ transforms as $U\rightarrow g_L U g_R^\dagger$ 
with $g_{L,R} \in G_{\rm global}$. 
There is an arbitrariness or a gauge degree of freedom (HLS) in dividing  
$U(x)$ into a product of  $\xi_L^\dagger$ and $\xi_R$ in such a way that they transform as 
 $\xi_{L,R} \rightarrow h(x) \xi_{L,R} g_{L,R}^\dagger$ and do the HLS gauge fields as 
$V_\mu(x) \rightarrow h(x)V_\mu(x) h^\dagger(x) + i h(x) \partial_\mu  h^\dagger(x)$ with 
$h(x) \in H_{\rm local}$. 
After gauge fixing of HLS, $H_{\rm local}$, 
the direct sum of the $H_{\rm local}$ and the subgroup $H_{\rm global}(\in G_{\rm global})$ 
becomes $H$ of the usual nonlinear sigma model manifold $G/H$. 
As done in the nonlinear sigma model, 
we may freely gauge $G_{\rm global}$ by introducing the external gauge fields ${\cal L}_\mu(x)$ 
and ${\cal R}_\mu(x)$ including the standard model
 gauge bosons $W,Z,\gamma$ through the standard promotion $g_{L,R} \Rightarrow g_{L,R}(x)$.

The HLS action $\Gamma^{\rm inv}_{\rm HLS} $ 
is thus constructed from invariant terms under 
parity ($P$) and charge conjugation ($C$) with IP-odd property~\footnote{
The intrinsic parity of a particle is assigned to be even, if its parity equals $(-1)^{\rm spin}$,  
and odd otherwise.  
} as well as the gauge transformation:  
\begin{eqnarray}
&& 
\delta_{\rm HLS} \Gamma^{\rm inv}_{\rm HLS} =0\,, 
\nonumber \\
\delta_{\rm HLS}: && 
\xi_{L,R} \rightarrow h(x) \xi_{L,R} g_{L,R}^\dagger(x) 
\,, \nonumber \\ 
&& 
{\cal L} \rightarrow g_{L} (x) {\cal L} g_{L}^\dagger(x) + i g_{L}(x) d g_{L}^\dagger (x)
\,, \nonumber \\ 
&& 
{\cal R} \rightarrow g_{R} (x) {\cal R} g_{R}^\dagger(x) + i g_{R}(x) d g_{R}^\dagger (x)
\,, \nonumber \\ 
&& 
V \rightarrow  h(x) V h^\dagger (x) + i h(x) d h^\dagger(x)
\,, \label{GT}
\end{eqnarray} 
where the differential form has been introduced: 
$d= (\partial_\mu) dx^\mu$, $V=V_\mu d x^\mu$, and so on. 
 The explicit expression of $\Gamma^{\rm inv}_{\rm HLS}$ 
takes the form~\cite{Fujiwara:1984mp,Bando:1987br,Harada:2003jx}: 
\begin{eqnarray} 
  \Gamma^{\rm inv}_{\rm HLS}
  [V, {\mathcal L} , {\mathcal R} , \xi_L^\dag \xi_R]
  &=&\frac{N_c}{16\pi^2} \int_{M^4} \sum_{i=1}^{4} c_i {\mathcal L}_i 
\,,   \label{sol:HLS}  \\ 
{\cal L}_1 
&=&
i {\rm tr} 
  [\hat{\alpha}_L^3 \hat{\alpha}_R - \hat{\alpha}_R^3 \hat{\alpha}_L ] 
\,, \\ 
{\cal L}_2 &=& 
i {\rm tr}
  [\hat{\alpha}_L \hat{\alpha}_R \hat{\alpha}_L \hat{\alpha}_R ] 
\,, \\ 
{\cal L}_3 &=&  
{\rm tr}
  [F_V(\hat{\alpha}_L \hat{\alpha}_R 
    - \hat{\alpha}_R \hat{\alpha}_L) ] 
\,, \\ 
{\cal L}_4 &=&  
\frac{1}{2} {\rm tr}
  [\hat{\cal F}_L (\hat{\alpha}_L \hat{\alpha}_R 
    - \hat{\alpha}_R \hat{\alpha}_L) 
- \hat{\cal F}_R (\hat{\alpha}_R \hat{\alpha}_L - 
   \hat{\alpha}_L \hat{\alpha}_R)] 
 \,, 
\end{eqnarray}
where the normalization of $c_1$-$c_4$ terms followed 
Ref.~\cite{Harada:2003jx} 
and 
\begin{eqnarray} 
\hat{\alpha}_{L} 
&=& 
  \frac{1}{i} d \xi_{L} \xi_{L}^\dagger - V 
  + \xi_{L} {\cal L} \xi_{L}^\dagger
\,, \qquad 
%\nonumber\\
\hat{\alpha}_{R} 
= 
  \frac{1}{i} d \xi_{R} \xi_{R}^\dagger - V 
  + \xi_{R} {\cal R} \xi_{R}^\dagger
\,, 
 \nonumber\\
\hat{\cal F}_{L} 
&=& \xi_{L}  {\cal F}_{L} \xi_{L}^\dagger 
\ , \qquad 
\hat{\cal F}_{R} 
= \xi_{R}  {\cal F}_{R} \xi_{R}^\dagger 
\,, \nonumber\\
{\cal F}_{L} 
&=& d {\cal L} - i {\cal L}^2
\,, \qquad 
{\cal F}_{R} 
= d {\cal R} - i {\cal R}^2
\,, 
 \nonumber\\
 F_V &=& d V - i V^2
\,.  
\end{eqnarray}   
It is evident that $\Gamma^{\rm inv}_{\rm HLS}$ is invariant under the gauge transformation Eq.(\ref{GT}).

Note that the $\omega$-$Z$-$\gamma$ vertex is not 
contained in the anomalous term $\Gamma_{\rm WZW}[{\cal L},{\cal R}, U]$ in Eq.(\ref{full:HLS}), but only exists in the gauge 
invariant terms  $\Gamma^{\rm inv}_{\rm HLS}[V, {\cal L},{\cal R}, \xi_L^\dag \xi_R]$ 
given by Eq.(\ref{sol:HLS}) which is 
{\it not determined by the anomaly}.  
We now demonstrate that the $\omega$-$Z$-$\gamma$ vertex indeed arises from 
$\Gamma^{\rm inv}_{\rm HLS}$.

\begin{figure} 

\begin{center} 
\includegraphics[scale=0.6]{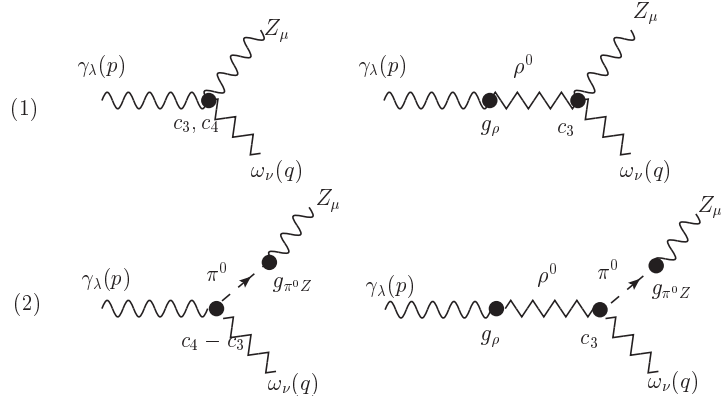}
\end{center}
\caption{ The diagrams relevant to the $\omega$-$Z$-$\gamma$ vertex (\ref{amp:totwZg}). }
\label{wZg-graphs}
\end{figure}

We employ the case with $N_f=2$ in which 
the two lightest quarks $(u,d)$ externally couple to the gauge fields ${\mathcal L} $ and ${\mathcal R} $. 
The ${\mathcal L} $ and ${\mathcal R} $ are then parametrized as   
\begin{equation} 
{\mathcal L} |_{\rm neutral} = e Q A + \frac{e}{s c} (T^3 - s^2 Q) Z 
\,, \qquad 
{\mathcal R}  |_{\rm neutral} = e Q \left(A - \frac{s}{c}  Z \right) 
\,, 
\end{equation}
where we have focused only on neutral gauge boson fields ($Z$ boson and photon ($A$) fields) 
which are relevant to the present study; $s \equiv \sin\theta_W$ is 
the weak mixing angle ($c^2=1-s^2$); $e$ the electromagnetic coupling constant; 
$Q$ the electric charge matrix $Q={\rm diag}(2/3, -1/3)$ and $T^3$ the isospin matrix for 
$(u,d)$ $T^3 = {\rm diag}(1/2, -1/2)$. 
The vector meson fields ($\rho^{\pm,0}, \omega$) 
are embedded in the HLS gauge field $V$ with the gauge couplings $g$ for $\rho^{\pm,0}$ and $g'$ for $\omega$, 
in such a way that 
\begin{equation}
  V = \frac{g}{2} \left( 
  \begin{array}{cc} 
  \rho^0 & \sqrt{2} \rho^+  \\ 
  \sqrt{2} \rho^- & - \rho^0   \\ 
\end{array}
\right) 
+ \frac{g'}{2} 
\left( 
  \begin{array}{cc} 
  \omega & 0 \\ 
  0 & \omega   \\ 
\end{array}
\right) 
\,. \label{V}
\end{equation}
 Note that $g'=g$ for $U(3)_L \times U(3)_R$ case.

  The  $\omega$-$Z$-$\gamma$ vertex function 
is thus constructed from 
  diagrams depicted in Fig.~\ref{wZg-graphs} as 
\begin{eqnarray} 
 \Gamma^{\mu\nu\lambda} [q, p+q, p] 
\Bigg|_{\omega Z\gamma} 
&=& 
\frac{N_c}{32\pi^2} \frac{e^2}{sc}g' 
\left( 
  \Gamma^{\mu\nu\lambda}_{(1)} [q, p+q, p] 
  + 
    \Gamma^{\mu\nu\lambda}_{(2)} [q, p+q, p] 
    \right) 
    \,, 
\label{amp:totwZg} \\ 
  \Gamma^{\mu\nu\lambda}_{(1)} [q, p+q, p] 
  &=& 
(c_3 + c_4) \, \epsilon^{\mu\nu\lambda\rho} p_\rho 
+ c_3 
\, \epsilon^{\mu\nu \rho \lambda'} (q-p)_\rho 
D_\rho(p^2) \left( 
 \frac{p^2 \delta^\lambda_{\lambda'} - p_{\lambda'} p^\lambda}{m_\rho^2} 
 \right) 
\,, \label{G1} \\ 
    \Gamma^{\mu\nu\lambda}_{(2)} [q, p+q, p] 
&=& \frac{(p+q)^\mu}{(p+q)^2} \epsilon^{\nu\lambda \alpha \beta} p_\alpha q_\beta 
\left[ 
 (c_4 - c_3) + 2 c_3 D_\rho(p^2) 
\right]  
\,, \label{G2}
\end{eqnarray} 
where we have read off the $\rho$ meson propagator $
 D_\rho(p^2) = m_\rho^2/(m_\rho^2 - p^2) $, 
the $\rho^0$-$\gamma$ and the $\pi^0$-$Z$ 
mixing strengths,   
$g_\rho = m_\rho^2/g $ and 
$g_{\pi^0 Z}=e/(2sc) F_\pi$, 
 from the IP-even sector~\cite{Harada:2003jx}. 
We thus conclude that the $\omega$-$Z$-$\gamma$ vertex includes free  
parameters $c_3$ and $c_4$ which are not determined by the anomaly 
in contrast to the claim of HHH~\cite{Harvey:2007rd,Harvey:2007ca}.  
In the next section  
we show that the $\omega$-$Z$-$\gamma$ vertex 
in the HHH formulation~\cite{Harvey:2007rd} also 
arises from the anomaly free (gauge invariant) terms 
having free parameters which are not determined by the anomaly.

Although the $\omega$-$Z$-$\gamma$ vertex generically includes the undetermined 
parameters $c_3$ and $c_4$,  it turns out that the form can be fixed by using 
phenomenological inputs associated with $\omega$-$\pi^0$-$\gamma$ vertex: 
This is possible because of the fact that the $\omega$-$Z$-$\gamma$ vertex 
is related to the $\omega$-$\pi^0$-$\gamma$ vertex   
by the chiral symmetry through the Ward-Takahashi identity,  
\begin{eqnarray} 
&&  k_\nu \,  \Gamma^{\mu \nu \lambda}[p+k, k, p] \Bigg|_{\omega Z \gamma} 
= k_\nu \left( 
\Gamma^{\mu \nu \lambda}_{(1)}[p+k, k, p] + \Gamma^{\mu \nu \lambda}_{(2)}[p+k, k, p] 
\right) 
= 0 
  \,,  \label{WT:full}
\end{eqnarray}   
which reads 
\begin{eqnarray} 
  k_\nu \,  \Gamma^{\mu \nu \lambda}_{(1)}[p+k, k, p] \Bigg|_{\omega Z \gamma}
  = 
- k_\nu \,  \Gamma^{\mu \nu \lambda}_{(2)}[p+k, k, p] \Bigg|_{\omega Z \gamma} 
  = \frac{e}{2sc} F_\pi \, \Gamma^{\mu \lambda}[p+k,k, p] \Bigg|_{\omega \pi^0 \gamma}
  \,. 
\end{eqnarray}
The amplitudes concerning the $\omega$-$Z$-$\gamma$ vertex 
can thus be free from $c_3$ and $c_4$ to be fixed 
by using experimental values associated with 
the $\omega$-$\pi^0$-$\gamma$ process, as we will see more explicitly later.

\section{The general solution in the HHH formulation} 
\label{comp-HHH}

We have shown that the $\omega$-$Z$-$\gamma$ vertex comes from the anomaly-free term in the HLS formulation. As we discussed in the Introduction, 
the same should be the case also in the framework that HHH was based on.    
In this section, in order to make the argument more concrete, we present the explicit form of  the general solution,  
$\Gamma_{\rm G-HHH}^{\rm inv}[{\cal L}, {\cal R}, B_L, B_R, U]$,   
to the homogeneous part of the WZ anomaly equation, Eq.(\ref{HHH:inv:eq}), 
including ``background fields" $B_L$ and $B_R$ introduced in 
Refs.~\cite{Harvey:2007rd,Harvey:2007ca}.  
It is shown that the HHH action $\Delta \Gamma_{\rm HHH}$ defined in Eq.(\ref{HHH:WZW})
is a particular choice of the general solution, 
so is the expression of the $\omega$-$Z$-$\gamma$ vertex given by 
HHH~\cite{Harvey:2007rd,Harvey:2007ca}: 
\begin{equation} 
 S_{\omega Z \gamma} \Bigg|_{\rm HHH} 
= 
 \frac{N_c}{16 \pi^2} \frac{e^2}{sc} g' 
\int_{M^4} \omega Z d A 
\,. \label{HHH:wZg}
\end{equation} 
The explicit comparison of the HHH action 
with the HLS action $\Gamma^{\rm inv}_{\rm HLS}$ in Eq.(\ref{sol:HLS}) 
will be given in Appendix~\ref{relation}.

The general solution to Eq.(\ref{HHH:inv:eq}) is constructed from the chiral (gauge) 
invariant terms having $P$- and $C$-even but IP-odd properties. 
The building blocks are classified into two pieces: 
variables transforming homogeneously with respect to either $g_L(x)$ or $g_R(x)$. 
  The possible set of the covariant variables is as follows: 
\begin{eqnarray} 
 {\cal O}_L &=& \{ B_L, U B_R U^\dag, {\cal D}B_L, {\cal F}_L, {\cal D}UU^\dag \} 
 \, \qquad 
  {\cal O}_L \to g_L(x) {\cal O}_L g_L^\dag(x) 
  \,, \nonumber \\ 
   {\cal O}_R &=& \{ B_R, U^\dag B_L U, {\cal D}B_R, {\cal F}_R, {\cal D}U^\dag U \} 
 \, \qquad 
  {\cal O}_R \to g_R(x) {\cal O}_R g_R^\dag(x) 
\,, \label{list}
\end{eqnarray} 
where 
 \begin{eqnarray} 
{\cal D}U &=& dU - i {\cal L} U + i U{\cal R} 
\,, \\ 
{\cal D}B_{L} &=& 
d B_{L} - i ({\cal L}  B_{L} + B_{L} {\cal L} )
\,, \\ 
{\cal D}B_{R} &=& 
d B_{R} - i ({\cal R}  B_{R} + B_{R} {\cal R}) 
\,. 
\end{eqnarray} 
With these at hand, we can find the general solution to Eq.(\ref{HHH:inv:eq}), 
\begin{eqnarray} 
\Gamma_{\rm G-HHH}^{\rm inv}[{\mathcal L} , {\mathcal R} , B_L,B_R,U]   
&=&  \frac{N_c}{16\pi^2}\int_{M^4} 
\sum_{i=1}^{14} a_i {\cal O}_i
\,, \label{general:HHH}
\end{eqnarray} 
which consists of a linear combination of the following fourteen terms~\footnote{
Without $a_1$ meson field, the number of the independent operators ${\cal O}_i$ is reduced to four 
as in the HLS formalism (See Appendix~\ref{relation}).  
Similar four terms were introduced in Ref.~\cite{Kaymakcalan:1984bz} based on the massive Yang-Mills approach 
for vector meson field. 
On the other hand, if the external fields are turned off in Eq.(\ref{general:HHH}) keeping the degree of freedom of $a_1$ meson,  
one would be left with eleven terms.  
In Ref.~\cite{Duan:2000dy} similar eleven terms were discussed based on the massive Yang-Mills approach. 
}: 
\begin{eqnarray} 
%({\cal L}_{1})_{\rm HHH}^{\rm inv} 
{\cal O}_1 
&=&
%a_1 
i \, {\rm tr}[ B_L^3 U B_R U^\dag - B_R^3 U^\dag B_L U ]  
\,, \nonumber \\ 
%({\cal L}_{2})_{\rm HHH}^{\rm inv} 
{\cal O}_2
&=&
%a_2 
i \, {\rm tr}[ B_L U B_R U^\dag B_L U B_R U^\dag  ] 
\,,  \nonumber \\ 
{\cal O}_3 
%({\cal L}_{3})_{\rm HHH}^{\rm inv} 
&=&
%a_3 
{\rm tr}[ ({\cal D} B_L B_L + B_L {\cal D} B_L)U B_R U^\dag  - ({\cal D} B_R B_R + B_R {\cal D} B_R)U^\dag B_L U ] 
\,, \nonumber \\ 
%({\cal L}_{4})_{\rm HHH}^{\rm inv} 
{\cal O}_4 
&=&
%a_4 
{\rm tr}[  ({\cal F}_L B_L + B_L {\cal F}_L) U B_R U^\dag - ({\cal F}_R B_R + B_R {\cal F}_R) U^\dag B_L U ] 
\,,  \nonumber \\ 
%({\cal L}_{5})_{\rm HHH}^{\rm inv} 
{\cal O}_5 
&=&
%a_5 
{\rm tr}[ B_L^3 ({\cal D}UU^\dag) - B_R^3 ({\cal D}U^\dag U) ] 
\,,  \nonumber \\ 
%({\cal L}_{6})_{\rm HHH}^{\rm inv} 
{\cal O}_6
&=&
%a_6 
{\rm tr} [ 
(B_L^2 U B_RU^\dag + U B_R U^\dag B_L^2) {\cal D}UU^\dag - (B_R^2 U^\dag B_L U + U^\dag B_L UB_R^2 ) {\cal D}U^\dag U ] 
\,,  \nonumber \\ 
%({\cal L}_{7})_{\rm HHH}^{\rm inv} 
{\cal O}_7 
&=&
%a_7 
{\rm tr}[ B_L U B_R U^\dag B_L ({\cal D}UU^\dag) - B_R U^\dag B_L U B_R ({\cal D}U^\dag U) ] 
\,,  \nonumber \\ 
%({\cal L}_{8})_{\rm HHH}^{\rm inv} 
{\cal O}_8
&=&
%a_8 
 i \, {\rm tr} [ ({\cal D}B_L B_L + B_L {\cal D}B_L) {\cal D}UU^\dag - ({\cal D}B_R  B_R + B_R {\cal D}B_R ) {\cal D}U^\dag U ] 
\,,  \nonumber \\ 
%({\cal L}_{9})_{\rm HHH}^{\rm inv} 
{\cal O}_9 
&=&
%a_9 
i \, {\rm tr} [ ({\cal D}B_L U B_R U^\dag + U B_R U^\dag {\cal D}B_L) {\cal D}UU^\dag - ({\cal D}B_R U^\dag B_L U + U^\dag B_L U {\cal D}B_R ) {\cal D}U^\dag U ] 
\,,  \nonumber \\ 
%({\cal L}_{10})_{\rm HHH}^{\rm inv} 
{\cal O}_{10}
&=&
%a_{10}  
i \, {\rm tr}[ ({\cal F}_L B_L + B_L {\cal F}_L) {\cal D}UU^\dag - ({\cal F}_R  B_R + B_R {\cal F}_R ) {\cal D}U^\dag U ] 
\,,  \nonumber \\ 
%({\cal L}_{11})_{\rm HHH}^{\rm inv} 
{\cal O}_{11}
&=&
%a_{11} 
i \, {\rm tr}[ ({\cal F}_L U B_R U^\dag + U B_R U^\dag {\cal F}_L) {\cal D}UU^\dag - ({\cal F}_R U^\dag B_L U + U^\dag B_L U {\cal F}_R ) {\cal D}U^\dag U ] 
\,,  \nonumber \\ 
%({\cal L}_{12})_{\rm HHH}^{\rm inv} 
{\cal O}_{12}
&=&
%a_{12} 
i \, {\rm tr} [ B_L U B_R U^\dag ({\cal D}UU^\dag)^2 - B_R U^\dag B_L U ({\cal D}U^\dag U)^2 ] 
\,,  \nonumber \\ 
%({\cal L}_{13})_{\rm HHH}^{\rm inv} 
{\cal O}_{13}
&=&
%a_{13} 
i \, {\rm tr} [ (B_L {\cal D}UU^\dag)^2 - (B_R {\cal D}U^\dag U)^2 ] 
\,,  \nonumber \\ 
%({\cal L}_{14})_{\rm HHH}^{\rm inv} 
{\cal O}_{14}
&=&
%a_{14} 
{\rm tr} [B_L ({\cal D} UU^\dag)^3 - B_R ({\cal D}U^\dag U)^3 ]  
\,. 
\end{eqnarray}
Because the terms in Eq.(\ref{general:HHH}) independently satisfy the homogeneous part of the WZ anomaly 
equation, Eq.(\ref{HHH:inv:eq}), the coefficients $a_1,a_2,\cdots , a_{14}$ cannot be fixed by the anomaly and hence 
should be treated as free parameters. 
This is in sharp contrast to the claim given by HHH~\cite{Harvey:2007ca} where the definite expression of the HHH action is 
given without any free parameters.    
It turns out that the HHH action actually corresponds to a  
particular choice for the free parameters $a_1,a_2,\cdots , a_{14}$, 
which is read off from Ref.~\cite{Harvey:2007ca} as 
\begin{eqnarray} 
&&  
a_1 = \frac{1}{3} 
\,, 
\qquad 
a_2 =  \frac{1}{6}
\,, 
\qquad 
a_3 =  - \frac{1}{3} 
\,, 
\qquad 
a_4 = - \frac{1}{3}
\,, \nonumber \\ 
&&  
a_5 =   \frac{1}{3}
\,, 
\qquad 
a_6 = 0 
\,, 
\qquad 
a_7 =  \frac{1}{3}
\,, 
\qquad 
a_8 =  \frac{1}{3}
\,, \nonumber 
\\ 
&&  
a_9 =  \frac{1}{6}
\,, 
\qquad 
a_{10} =  \frac{2}{3}
\,, 
\qquad 
a_{11} =  \frac{1}{3}
\,, 
\qquad 
a_{12} =  \frac{1}{3}
\,, \nonumber 
\\ 
&&  
a_{13} = - \frac{1}{6}
\,, 
\qquad 
a_{14} =  \frac{1}{3}
\,. \label{a14:val} 
\end{eqnarray}
We may therefore express the HHH action $\Delta \Gamma_{\rm HHH}$ defined in Eq.(\ref{HHH:WZW}) 
as  
\begin{equation} 
  \Delta \Gamma_{\rm HHH} = \Gamma_{\rm G-HHH}^{\rm inv} \quad {\rm with} \quad 
a_{1-14} \quad {\rm in} \quad {\rm Eq.(\ref{a14:val})} 
\,. \label{delta:HHH} 
\end{equation}

By putting the $\omega$ meson field into $B_L$ and $B_R$ as done in Ref.~\cite{Harvey:2007ca} 
\begin{equation} 
B_L + B_R = 
 g' 
\left( 
\begin{array}{cc} 
 \omega & 0 \\ 
 0 & \omega 
\end{array} 
\right) 
+ \cdots 
\,, 
\end{equation}
the $\omega$-$Z$-$\gamma$ vertex 
is read off from the general solution Eq.(\ref{general:HHH}): 
 \begin{equation} 
S_{\omega Z \gamma} 
= 
 \frac{N_c}{16 \pi^2} \frac{e^2}{sc} g' (a_{10} + a_{11}) 
\int_{M^4} \omega Z d A 
\,,   
\end{equation}
which is reduced to 
the HHH result~\cite{Harvey:2007ca} in Eq.(\ref{HHH:wZg})  for $a_{10} + a_{11}=1$ (See Eq.(\ref{a14:val})).

The expression of $\Gamma_{\rm G-HHH}^{\rm inv}$ in Eq.(\ref{general:HHH}) actually becomes 
identical to that obtained in the framework of the GHLS~\cite{Bando:1984pw,Bando:1987br,Bando:1987ym} 
which includes fourteen invariant terms as well~\cite{Kaiser:1990yf}. 
This can be seen just by converting the gauge fields $L$ and $R$ associated 
with the GHLS $G_{\rm local }=[U(N_f)_L \times U(N_f)_R]_{\rm local}$ into 
the background fields $B_L$ and $B_R$ through a certain operation as shown in Appendix~\ref{GHLS:inv}.

\section{Application to neutrino-nucleon collision processes}  
\label{implications}

In this section we shall address the phenomenological 
application of the $\omega$-$Z$-$\gamma$ vertex 
to neutrino ($\nu$) - nucleon ($N$) collision processes 
 obtained from the HLS formulation as well as the HHH formulation.

\begin{figure} 

\begin{center} 
\includegraphics[scale=0.6]{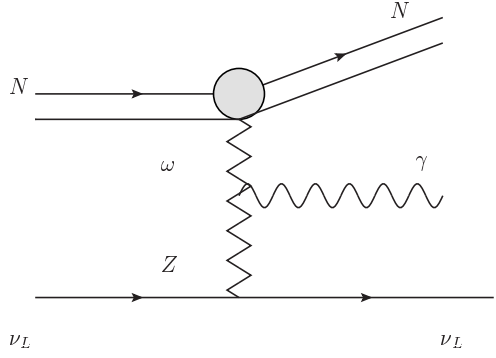}
\end{center}
\caption{ The $\nu N \to \nu N \gamma$ process arising from the $\omega$-$Z$-$\gamma$ vertex. } 
\label{Nnu-Nnug-graph}
\end{figure}

We first discuss contributions to a $\nu N \to \nu N \gamma$ process 
coming from the $\omega$-$Z$-$\gamma$ vertex as depicted in  Fig.~\ref{Nnu-Nnug-graph}. 
We consider the heavy nucleon limit so that the nucleon does not move to be completely stationary. 
In this limit the pion exchange contributions corresponding to terms in $\Gamma_{(2)}^{\mu\nu\lambda}$ of Eq.(\ref{amp:totwZg}) 
vanish in the amplitude. 
Integrating out the $Z$ boson and replacing it with the neutrino current 
$J_\mu= \bar{\nu}_L \gamma_\mu \nu_L$,  
we can then write the effective action relevant to this process as 
\begin{equation} 
  S_{\rm eff} = \kappa \int d^4 x d^4 y 
\epsilon^{0ijk}  
\delta^{(3)}(\vec{x})  \, {\cal F}_{\omega^* Z^*\gamma^*}(x-y) \,  
 J_i (y) F_{jk}(y) 
\,, \label{Seff:nuNnuNg}
\end{equation} 
 where $F_{jk}$ stands for the photon field strength. 
The $\omega^*$-$Z^*$-$\gamma^*$ form factor, the Fourier transformation of 
${\cal F}_{\omega^* Z^*\gamma^*}(x-y)$,  
 can be read off from the diagram shown in Fig.~\ref{Nnu-Nnug-graph}:  
\begin{equation} 
  {\cal F}_{\omega^* Z^*\gamma^*}(q^2) = 
 \frac{1-\bar{c}}{2} +  \frac{1+\bar{c}}{2} D_\rho(q^2) 
  \,, \label{F:wZg}
\end{equation}  
where 
\begin{eqnarray} 
  \bar{c}\,|_{\rm HLS} &=& \frac{c_3-c_4}{c_3+c_4}  
  \,, 
\label{ff:oZNN:HLS}
\end{eqnarray}
 for the HLS formulation (See Eq.(\ref{amp:totwZg})) and likewise 
\begin{eqnarray} 
  \bar{c} \, |_{\rm G-HHH} &=& \frac{2(a_8 + a_9)}{a_{10} + a_{11}} - 1   
  \,, 
\label{ff:oZNN:GHHH}
\end{eqnarray}
 for the HHH formulation. 
The overall coupling $\kappa$ is given by 
\begin{eqnarray} 
   \kappa\,|_{\rm HLS} &=& \frac{N_c}{8 \sqrt{2} \pi^2} \, 
  \frac{e g_\omega G_F}{m_\omega^2} \, 
\frac{g'(c_3+c_4)}{2} 
  \, , \\
    \kappa \, |_{\rm G-HHH} &=& \frac{N_c}{8 \sqrt{2} \pi^2} \, 
  \frac{e g_\omega G_F}{m_\omega^2} \, g'(a_{10}+a_{11}) 
  \, . 
\end{eqnarray}
Here   
$g_\omega$ is a coupling strength between the nucleon current $J^N_\mu = \bar{N} \gamma_\mu N$ and the $\omega$ meson  
defined by ${\cal L}_{\omega NN} = g_\omega \omega^\mu J_\mu^N   $.  
 From the effective action (\ref{Seff:nuNnuNg}),  
we compute the total cross section $\sigma(\nu N \to \nu N \gamma)$ as a function of 
the incident neutrino energy $E_\nu$ evaluated at the rest frame of nucleon, with nucleon recoil being neglected: 
 For the HLS formulation, we obtain    
\begin{eqnarray} 
 \sigma(\nu N \to \nu N \gamma)  \Bigg|_{\rm HLS}
& =& \frac{3 \alpha g_\omega^2 G_F^2}{640 \pi^6 m_\omega^4} \left(\frac{g' (c_3+c_4)}{2} \right)^2 E_\nu^6 
\,, \label{sigma:HLS} 
\end{eqnarray} 
and similarly for the HHH formulation: 
\begin{eqnarray} 
 \sigma(\nu N \to \nu N \gamma)  \Bigg|_{\rm G-HHH}
& =& \frac{3 \alpha g_\omega^2 G_F^2}{640 \pi^6 m_\omega^4} g'^2 (a_{10} + a_{11})^2 E_\nu^6 
\,. \label{sigma:GHHH} 
\end{eqnarray} 
The cross section $\sigma(\nu N \to \nu N \gamma) $ is thus determined by 
the free parameters $c_3$ and $c_4$ ($a_{10}$ and $a_{11}$) not fixed by the anomaly, 
in contrast to the claim of HHH~\cite{Harvey:2007rd}:  
The HHH result~\cite{Harvey:2007rd} is nothing but 
a particular choice taking $a_{10} + a_{11}=1$ (See Eq.(\ref{a14:val})): 
\begin{eqnarray} 
 \sigma(\nu N \to \nu N \gamma)  \Bigg|_{\rm HHH}
& =& \frac{3 \alpha g_\omega^2 G_F^2}{640 \pi^6 m_\omega^4} g'^2  E_\nu^6 
\,. \label{sigma:HHH} 
\end{eqnarray}

Although the cross section $\sigma(\nu N \to \nu N \gamma)$ includes the free parameters, 
it turns out that it can be fixed by using 
phenomenological inputs associated with $\omega$-$\pi^0$-$\gamma$ vertex.  
This is possible due to the Ward-Takahashi identity (\ref{WT:full}) associated with the chiral symmetry regarding 
the $Z$-boson current: 
Setting the $\omega$ momentum to zero $p+k=0$ in Eq.(\ref{WT:full}), which corresponds to 
our process, we have 
 \begin{eqnarray} 
  k_\nu \,  \Gamma^{\mu \nu \lambda}_{(1)}[0, k, -k] \Bigg|_{\omega Z \gamma}
  = \frac{e}{2sc} F_\pi \, \Gamma^{\mu \lambda}[0,k, -k] \Bigg|_{\omega \pi^0 \gamma}
  \,. \label{WT}
\end{eqnarray}
This implies that   
the cross section $\sigma(\nu N \to \nu N \gamma)$ is expressed 
by using the $\omega \to \pi^0 \gamma$ decay width $\Gamma(\omega \to \pi^0 \gamma)$: 
In the HLS formulation we have~\cite{Harada:2003jx}  
 \begin{equation} 
  \Gamma(\omega \to \pi^0 \gamma) \Bigg|_{\rm HLS}
  = \frac{3 \alpha}{64\pi^4 F_\pi^2} \left( \frac{m_\omega^2 - m_{\pi^0}^2}{2 m_\omega} \right)^3 \left(  \frac{g'(c_3+c_4)}{2}  \right)^2 
\,. \label{Gamma:HLS}
\end{equation}
Likewise in the HHH formulation, we have  
 \begin{equation} 
  \Gamma(\omega \to \pi^0 \gamma) \Bigg|_{\rm G-HHH}
  = \frac{3 \alpha}{64\pi^4 F_\pi^2} \left( \frac{m_\omega^2 - m_{\pi^0}^2}{2 m_\omega} \right)^3 g'^2 (a_{10} + a_{11})^2 
\,. \label{Gamma:GHHH}
\end{equation}
Although $\Gamma(\omega \to \pi^0 \gamma)|_{\rm HLS} $ and $\Gamma(\omega \to \pi^0 \gamma)|_{\rm G-HHH} $ 
 have free parameters ($c_3$, $c_4$) in the HLS formulation and 
($a_{10}$, $a_{11}$) in the HHH formulation, respectively, 
the ratio  
\begin{equation} 
  \frac{\sigma (\nu N \to \nu N \gamma) }{\Gamma (\omega \to \pi^0 \gamma)}  
  = \frac{g_\omega^2}{10 \pi^2} \left( \frac{2 m_\omega}{m_\omega^2 - m_{\pi^0}^2} \right)^3 
\frac{G_F^2 F_\pi^2 E_\nu^6}{m_\omega^4}
\, \label{HLSfree-ratio}
\end{equation}
is free from the parameters and hence is fixed (up to $g_\omega$)
once the experimental value of $\Gamma (\omega \to \pi^0 \gamma)$ is used as 
input.  
We thus evaluate the cross section to get  
\begin{equation} 
 \sigma (\nu N \to \nu N \gamma)  
 = 3.0 \times 10^{-41} \, {\rm cm}^2  \,
\left( \frac{\Gamma (\omega \to \pi^0 \gamma)}{0.70 \,{\rm MeV}} \right) 
\left( \frac{g_\omega}{13.4} \right)^2 
\left( \frac{E_\nu}{\rm GeV} \right)^6 
\,, \label{HLS:prediction}
\end{equation} 
where use has been made of the reference values $g_{\omega} \simeq 13.4$~\cite{Kapusta} 
and $\Gamma (\omega \to \pi^0 \gamma) = 0.70 \pm 0.03$ MeV~\cite{Nakamura:2010zzi}. 
As was indicated in Ref.~\cite{Harvey:2007rd}, this cross section may explain 
the excess of electron-like events in a low-energy range of the  
the quasi-elastic (QE) $\nu$-$N$ collision process, 
$200 \, {\rm MeV} \lesssim E_\nu^{\rm QE} \lesssim 475\, {\rm MeV}$,  
which has recently been observed 
at the Fermilab Booster Neutrino Experiment MiniBooNE~\cite{AguilarArevalo:2008rc,Djurcic:2009ds}, 
where electrons could be mimicked by a hard photon $\gamma$ 
at the detector.

\begin{figure}
\begin{center}
\includegraphics[scale=1.0]{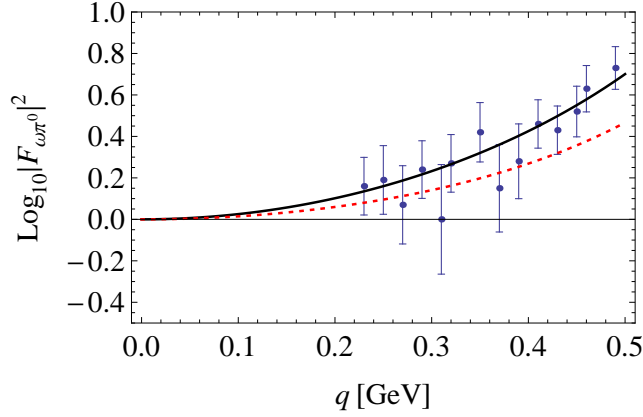}
\vspace{10pt} 
\caption{The best fit curve of the $\omega$-$\pi^0$ transition 
form factor $F_{\omega \pi^0}(q^2)$ yielding  
$\chi^2/{\rm d.o.f}=4.3/13=0.33$ with $\bar{c}=0.74$ (black solid line) 
compared with the case of the $\rho$ meson dominance with $\bar{c}=0$ (red dotted line: $\chi^2/{\rm d.o.f}=23/14 = 1.6$) 
together with data from the NA60 experiment~\cite{:2009wb}. } 
\label{Fwpi-fit} 
\end{center}
\end{figure}

We shall next present a prediction to $\nu + N \to \nu + N + \gamma^*$ process 
where $\gamma^*$ can be a charged lepton pair $l^\pm$.   
 From the effective action (\ref{Seff:nuNnuNg}), 
we straightforwardly evaluate the cross section at the rest-frame of the nucleon 
with the incident neutrino energy $E_\nu$ and nucleon recoil being neglected, 
to get  
\begin{eqnarray} 
 \sigma (\nu N \to \nu N \gamma^*(q^2)) 
& =& \sigma (\nu N \to \nu N \gamma) 
\cdot f(q^2)  
\,, \label{sigma2:HLS} 
\end{eqnarray}  
 where $\sigma (\nu N \to \nu N \gamma) $ is given in Eq.(\ref{sigma:HLS}) and 
\begin{eqnarray} 
f(q^2) &=&  
|{\cal F}_{\omega^*Z^*\gamma^*}(q^2)|^2 \cdot 
\Bigg[ 
  \left\{ 
   1 + \frac{133}{16} \frac{q^2}{E_\nu^2} - \frac{643}{32} \left( \frac{q^2}{E_\nu^2}  \right)^2 
   \right\} \sqrt{1 - \frac{q^2}{E_\nu^2}}  
  \nonumber \\ 
  && 
  + \frac{15}{4} \left\{  6 + \frac{q^2}{E_\nu^2} \right\} \left( \frac{q^2}{E_\nu^2} \right)^2 
  \ln \left( \frac{E_\nu + \sqrt{E_\nu^2- q^2}}{\sqrt{q^2}}  \right) 
\Bigg] 
\,, \label{fq2} 
\end{eqnarray} 
and ${\cal F}_{\omega^* Z^*\gamma^*}(q^2)$ is given in Eq.(\ref{F:wZg}). 
In order to evaluate Eq.(\ref{sigma2:HLS}) explicitly, 
the free parameter $\bar c$ in Eq.(\ref{F:wZg}) needs to be fixed. 
It turns out that the $\bar c$ can be determined 
 by fitting the $\omega$-$\pi^0$ transition form 
factor to the $\omega \to \pi^0 l^+l^-$ decay data~\cite{:2009wb}:  
Consider the $\omega \to \pi^0 l^+ l^-$ decay width
\begin{eqnarray} 
\Gamma(\omega \to \pi^0 l^+l^-) 
&=& 
  \int^{(m_\omega - m_{\pi^0})^2}_{4 m_l^2} dq^2 
  \frac{\alpha}{3\pi} \frac{\Gamma(\omega \to \pi^0 \gamma)}{q^2}
  \left(  1+ \frac{2 m_l^2}{q^2} \right) \sqrt{\frac{q^2-4 m_l^2}{q^2}} 
\nonumber \\ 
&& \times \left[ 
 \left(1 + \frac{q^2}{m_\omega^2-m_{\pi^0}^2} \right)^2 
 - \frac{4 m_\omega^2 q^2}{(m_\omega^2 - m_{\pi^0}^2)^2}
\right]^{3/2} \cdot |F_{\omega \pi^0} (q^2) |^2
\,,  
\end{eqnarray}  
where $F_{\omega \pi^0}$ denotes 
the $\omega$-$\pi^0$ transition form factor. 
  For $0 \le q^2 \lesssim 1\, {\rm GeV}^2$ 
the general form of $F_{\omega \pi^0}(q^2)$ normalized as $F_{\omega \pi^0}(0)=1$ 
is given by   
\begin{equation} 
  F_{\omega \pi^0}(q^2) 
  =  D_\rho(q^2) + \bar{c} \cdot [D_\rho(q^2)-1] 
  \,. \label{HLSFwpi}
\end{equation}
Note that $F_{\omega \pi^0}$ includes 
 the same parameter $\bar c$ as in  
${\cal F}_{\omega^* Z^*\gamma^*}$ of Eq.(\ref{F:wZg}),  
which reflects the fact that the chiral 
symmetry relates the $\omega$-$\pi^0$-$\gamma$ process with 
the $\omega$-$Z$-$\gamma$ process through the Ward-Takahashi identity in Eq.(\ref{WT}). 
Performing the fit to the experimental data on $F_{\omega \pi^0}$ measured at the NA60 experiment~\cite{:2009wb}, 
we find the best fit value of $\bar{c}$ (with $\chi^2/{\rm d.o.f}=4.3/13=0.33$), 
\begin{equation} 
  \bar{c} |_{\rm best}= 0.74
  \,. \label{cbar:best}
\end{equation}
The corresponding curve  of $F_{\omega \pi^0}$ is shown in Fig.~\ref{Fwpi-fit}.

In contrast, 
the parameter choice Eq.(\ref{a14:val}) corresponding to the HHH 
action~\cite{Harvey:2007ca} leads to ${\bar c}\, |_{\rm HHH}=0$ 
($\rho$ meson dominance) 
once a canonical kinetic term of $\rho$ is assumed, 
and hence 
does not achieve the best fit form of $F_{\omega \pi^0}$.

\begin{figure}[t]
\begin{center}
 \includegraphics[scale=1.0]{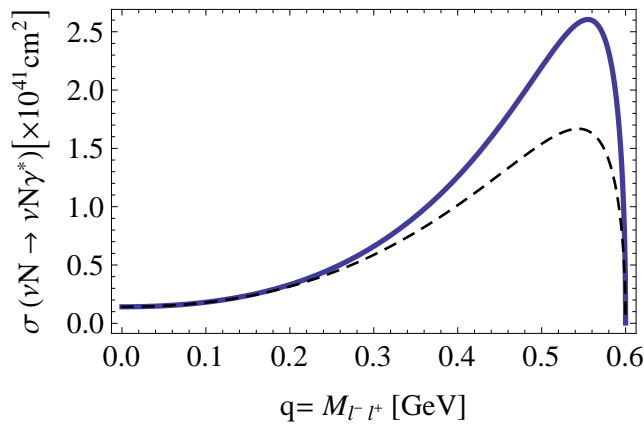} 
\vspace{10pt} 
\caption{ The total cross section $\sigma(\nu N \to \nu N \gamma^*)$ 
as a function of $q=M_{l^+l^-}$ 
with the incident neutrino energy $E_\nu = 600$ MeV fixed. 
The solid curve is drawn by using the best fit value of $\bar{c}$, 
$\bar{c}|_{\rm best} =0.74$, while the dotted curve corresponds to the $\rho$ meson dominance ($\bar{c}=0$). } 
\label{sigmaHLSfig} 

\end{center}
\end{figure}

Now, using the best fit value $\bar{c}|_{\rm best}=0.74$,  
we obtain Fig.~\ref{sigmaHLSfig} which shows 
the predicted curve of the total cross section $\sigma(\nu N\to \nu N \gamma^*)$ 
with $E_\nu = 600$ MeV fixed as a function of the virtual photon momentum $q$ 
which could be an invariant mass of  $l^\pm$ pair production, $q=M_{l^-l^+}$.  
Remarkably, the cross section $\sigma(\nu N \to \nu N \gamma^*)$ 
 has a peak around $500$ MeV which will yield somewhat larger number of events in 
this energy range~\footnote{ 
 Higher order terms of ${\cal O}(p^4)$ in derivative expansion such as the $z_3$ term  
would affect the $\rho$-$\gamma$ mixing strength $g_\rho$ by about 10\% 
when it is evaluated at the on-shell of $\rho$ meson~\cite{Harada:2003jx}. 
The definition of ${\bar c}$ in Eq.(\ref{ff:oZNN:HLS}) will then be modified 
involving the $z_3$-term contribution, but the form of Eq.(\ref{F:wZg}) turns out to be intact, 
so is the prediction about the cross section $\sigma(\nu N \to \nu N \gamma^*)$. }. 
This enhancement essentially comes from the dynamical $\rho$-contribution and is 
independently of the value of $g_\omega$, which is to be tested at $\nu$-$N$ 
collision experiments like the MiniBooNE in the future.

\section{Summary} 
\label{summary}

In this paper we presented the general homogeneous solution 
to the WZ anomaly equation in the HHH formulation, which was given by 
a linear combination of fourteen gauge 
invariant terms with the coefficients not determined by the anomaly.  
It was clarified that the definite form of the HHH action $\Delta \Gamma_{\rm HHH}$ 
 is nothing but a particular expression for 
the general solution $\Gamma_{\rm G-HHH}^{\rm inv}$ in Eq.(\ref{general:HHH}) (See Eq.(\ref{delta:HHH})).

We formulated the general form of $\omega$-$Z$-$\gamma$ vertex 
arising from the gauge invariant HLS action $\Gamma_{\rm HLS}^{\rm inv}$ 
in Eq.(\ref{sol:HLS}) 
as well as the G-HHH action $\Gamma_{\rm G-HHH}^{\rm inv}$ in Eq.(\ref{general:HHH}) having  
free parameters not determined by the anomaly.  
In spite of the free parameters, 
the $\omega$-$Z$-$\gamma$ vertex related to the 
$\nu$ -$N$ collision process 
was determined by using the experimental input for the $\omega \to \pi^0 \gamma$ decay width 
$\Gamma(\omega \to \pi^0 \gamma)$  through the Ward-Takahashi identity. 
Other set of the free parameters was also used to make the best fit to 
the experimental data on the $\omega$-$\pi^0$-$\gamma^*$ process, 
which provided us with a prediction to the cross section $\sigma(\nu N \to \nu N \gamma^*)$ 
to be tested at $\nu$-$N$ collision experiments in the future.

\section*{Acknowledgments}

We would like to thank Chris Hill and Richard Hill for stimulating discussions 
during the stay of K.Y. at Fermilab in 2008 summer when this work was initiated.  
This work was supported in part by 
the JSPS Grant-in-Aid for Scientific Research (S) \#22224003 
and the Grant-in-Aid for Nagoya University Global COE Program, ``Quest for Fundamental 
Principles in the Universe: from Particles to the Solar System and the Cosmos'', from the 
Ministry of Education, Culture, Sports, Science and Technology of Japan (M.H. and K.Y.).  
M.H. was supported in part by 
the JSPS Grant-in-Aid for Scientific Research (c) 20540262 and
Grant-in-Aid for Scientific Research on Innovative Areas (No. 2104) 
``Quest on New Hadrons with Variety of Flavors'' from MEXT. 
S.M. was supported by 
the U.S. Department of Energy under Grant
No. DE-FG02-06ER41418 and the Korea Research 
Foundation Grant funded by the Korean Government (KRF-2008-341-C00008).

\appendix 
\renewcommand\theequation{\Alph{section}.\arabic{equation}}

\section{The explicit relation between the HLS and HHH formulations} 
\label{relation}

In this Appendix we explicitly compare the HHH action Eq.(\ref{delta:HHH}) with the HLS action Eq.(\ref{sol:HLS}) by 
eliminating axialvector mesons out of the HHH action. 
 For that purpose, it is convenient to introduce ``matter fields" $B_V$ and $B_A$: 
\begin{equation} 
B_V =\frac{\xi(\pi) B_R \xi^\dag(\pi) + \xi^\dag(\pi) B_L \xi(\pi)}{2} 
\,, \qquad 
B_A =\frac{\xi(\pi) B_R \xi^\dag(\pi) - \xi^\dag(\pi) B_L \xi(\pi)}{2} 
\,, \label{BA}
\end{equation}
where $\xi(\pi)$ and $\xi^\dag(\pi)$ denote the representatives of coset space $G/H$.  
Since $\xi(\pi)$ and $\xi^\dag(\pi)$ transform under $G$ as 
$\{\xi(\pi), \xi^\dag(\pi)\} \to \{ h(\pi) \xi(\pi) g_L^\dag, g_R \xi(\pi) h^\dag(\pi) \}$ 
where $h(\pi) \in H$, $B_V$ and $B_A$ defined in Eq.(\ref{BA}) transform as 
\begin{equation} 
B_{V,A} \to h(\pi) B_{V,A} h^\dag (\pi) 
\,.\label{matter:trans}
\end{equation}
We also introduce a 1-form $\alpha_\perp(\pi)$ to replace ${\cal D}U$ terms in Eq.(\ref{delta:HHH}) 
by $\alpha_\perp(\pi) = \xi^\dag(\pi) {\cal D}U \xi^\dag(\pi)/(2i)$ which transforms 
in the same way as $B_{V,A}$: 
\begin{equation} 
\alpha_\perp(\pi) \to h(\pi) \alpha_\perp(\pi) h^\dag(\pi)
\,. \label{alphapi:trans}
\end{equation}   
We now remove axialvector mesons by putting $B_A$ in Eq.(\ref{BA}) to be zero 
and express all the remaining terms in Eq.(\ref{delta:HHH}) 
in terms of $B_V$ and $\alpha_\perp(\pi)$.  
This operation is equivalent to solving away the axialvector meson field 
through its equation of motion, just like a way~\cite{Harada:2010cn} of integrating out higher Kaluza-Klein modes 
arising in holographic QCD,     
which keeps the gauge invariance manifestly. 
We then find that the ${\cal O}_{1}$-${\cal O}_{4}$, ${\cal O}_{12}$ and ${\cal O}_{13}$  
terms in Eq.(\ref{delta:HHH}) vanish and the remaining terms are reduced to only the following four terms:  
\begin{eqnarray} 
\Delta{\Gamma}_{\rm HHH} [B_V, {\cal L}, {\cal R}, \xi^2(\pi)]
 &=& 
\frac{N_c}{16\pi^2} \int_{M^4} \Bigg[ 
a_1' \, i{\rm tr}[\alpha_\perp(\pi) B_V^3] 
+ a_2' \, i {\rm tr}[B_V \alpha_\perp^3(\pi)] 
 \nonumber \\ 
&& 
+ a_3' \, {\rm tr}[(\alpha_\perp(\pi) B_V - B_V \alpha_{\perp}(\pi)) {\cal D}B_V]  
+ a_4' \, {\rm tr} [(\alpha_\perp(\pi) B_V - B_V \alpha_{\perp}(\pi)) \hat{\cal F}_V(\pi)] 
\Bigg] 
\,, \label{general:HHH:redu}
\end{eqnarray}
with 
\begin{eqnarray} 
  a_1' &=&  -4 a_5 - 8 a_6 - 4 a_7 = 
- \frac{8}{3} 
\,, \nonumber \\ 
    a_2' &=& 16 a_{14} = 
- \frac{16}{3}
\,, \nonumber \\ 
    a_3' &=&  4 a_8 + 4 a_9 = 2 
\,, \nonumber \\ 
    a_4' &=&  4 a_{10} + 4 a_{11} = 4 
    \,, \label{a4:HHH}
\end{eqnarray} 
where 
\begin{eqnarray} 
{\cal D}B_V &=& d B_V - i (\alpha_{||}(\pi) B_V + B_V \alpha_{||}(\pi)) 
\,, \nonumber \\ 
\alpha_{||}(\pi) &=& 
\frac{1}{2i}\left[ 
d \xi(\pi) \xi^\dag(\pi) + d \xi^\dag(\pi) \xi(\pi) + i \xi(\pi) {\mathcal R}  \xi^\dag(\pi) + i \xi^\dag(\pi) {\mathcal L}  \xi(\pi)
\right]  
\,, \nonumber \\ 
\hat{\cal F}_V(\pi)&=&
\frac{1}{2} ( \xi(\pi) {\cal F}_R \xi^\dag(\pi) + \xi^\dag(\pi) {\cal F}_L \xi(\pi))  
\,.   
\end{eqnarray}

The HLS action $\Gamma^{\rm inv}_{\rm HLS}$ in Eq.(\ref{sol:HLS}) is, on the other hand, 
rewritten into the following form: 
\begin{eqnarray} 
 \Gamma^{\rm inv}_{\rm HLS}[V, {\cal L}, {\cal R}, \xi^\dag_L \xi_R] 
 &=& \frac{N_c}{16\pi^2} \int_{M^4} \Bigg[ 
 -4 (c_1+c_2) i {\rm tr}[\hat{\alpha}_\perp \hat{\alpha}_{||}^3] 
 + 4 (c_1-c_2) i {\rm tr}[\hat{\alpha}_{||} \hat{\alpha}_\perp^3] 
 \nonumber \\ 
&& 
- 2 c_3 {\rm tr}[(\hat{\alpha}_\perp \hat{\alpha}_{||} - \hat{\alpha}_{||} \hat{\alpha}_{\perp}) F_V] 
- 2 c_4 {\rm tr} [(\hat{\alpha}_\perp \hat{\alpha}_{||} - \hat{\alpha}_{||} \hat{\alpha}_{\perp}) \hat{\cal F}_V] 
\Bigg] 
\,, \label{wr:ipodd} 
\end{eqnarray}
where 
\begin{eqnarray} 
\hat{\alpha}_{||,\perp} &=& \frac{1}{2}(\hat{\alpha}_R \pm \hat{\alpha}_L) 
\,, \nonumber \\ 
\hat{\cal F}_V &=& \frac{1}{2} (\hat{\cal F}_R + \hat{\cal F}_L)  
\,. 
\end{eqnarray} 
Note that, in the unitary gauge of HLS ($\xi_L^\dag=\xi_R = \xi(\pi)$), 
$\hat{\alpha}_{||}$ transforms in the same way as the vector meson field $B_V$ in Eq.(\ref{general:HHH:redu}), 
$\hat{\alpha}_{||} \to h(\pi) \hat{\alpha}_{||} h^\dag(\pi)$~\cite{Harada:2003jx}, 
while $\hat{\alpha}_\perp$ becomes identical to $\alpha_\perp(\pi)$ in Eq.(\ref{general:HHH:redu}).  
We may therefore identify $\hat{\alpha}_{||}$ with the vector meson field $B_V$ in Eq.(\ref{general:HHH:redu}), 
in such a way that  $B_V=\eta \cdot \hat{\alpha}_{||}$ with a parameter $\eta$. 
In the unitary gauge, Eq.(\ref{wr:ipodd}) is thus expressed in terms of $B_V$ and $\alpha_\perp(\pi)$ to be 
\begin{eqnarray} 
 && 
\Gamma^{\rm inv}_{\rm HLS: \,unitary}[B_V, {\cal L}, {\cal R}, \xi^2(\pi)] 
\nonumber \\ 
&& = \frac{N_c}{16\pi^2} \int_{M^4} \Bigg[ 
 \left\{ \frac{-4 (c_1+c_2-c_3)}{\eta^3} \right\}  \cdot i{\rm tr}[\alpha_\perp(\pi) B_V^3] 
 + \left\{ \frac{4 (c_1-c_2 + c_3)}{\eta} \right\} \cdot i {\rm tr}[B_V \alpha_\perp^3(\pi)] 
 \nonumber \\ 
&& 
 + \left\{ \frac{2 c_3}{\eta^2} \right\} \cdot {\rm tr}[(\alpha_\perp(\pi) B_V - B_V \alpha_{\perp}(\pi)) {\cal D}B_V]  
+ \left\{ \frac{- 2 (c_3+c_4)}{\eta} \right\} \cdot {\rm tr} [(\alpha_\perp(\pi) B_V - B_V \alpha_{\perp}(\pi)) \hat{\cal F}_V(\pi)] 
\Bigg] 
\,,   
\label{B:ipodd}
\end{eqnarray} 
which is precisely the same form as the HHH action Eq.(\ref{general:HHH:redu}). 
  We thus find the particular choice for $c_1$-$c_4$, 
\begin{eqnarray}
c_1 + c_2 - c_3 &=& 
\frac{2}{3} \eta^3 
\,, \label{consc1} \\ 
c_1 - c_2 +c_3 &=&  
- \frac{4}{3} \eta 
\,, \label{consc2}\\ 
c_3 &=& 
\eta^2 
\,, \\ 
c_3 + c_4 &=& 
 -2 \eta 
\,, \label{consc4}
\end{eqnarray} 
which clarifies that the HHH action corresponds to a particular expression for  
$\Gamma^{\rm inv}_{\rm HLS}$, so does the $\omega$-$Z$-$\gamma$ vertex.

The physical implication of such a choice can be seen as follows: 
If we apply the parameter choice (\ref{consc1})-(\ref{consc4}) to the HLS formulation and 
impose the vector meson dominance on the $\pi^0 \rightarrow \gamma\gamma$ in accord with
the experiment data on the $\pi^0$-$\gamma$ transition form factor~\cite{Nakamura:2010zzi}, 
we would get $c_3+c_4=2$~\cite{Fujiwara:1984mp,Bando:1987br,Harada:2003jx},  which implies $\eta=-1$.
Then the above relations indicate that $c_3=c_4=1$, $c_1=1/3$, $c_2=0$.  
This is a special case of ``complete vector meson dominance'' 
 $c_3=c_4=1$, $c_1-c_2=1/3$,  which is known to be in contradiction with the experimental data on 
$\omega \to \pi^0 \pi^+ \pi^-$ (see for a recent argument Ref.~\cite{Harada:2003jx}).

\section{The G-HHH action and the IP-odd gauge invariant terms in the GHLS formalism } 
\label{GHLS:inv}

In this appendix we relate the G-HHH action $\Gamma_{\rm G-HHH}^{\rm inv}$ in Eq.(\ref{general:HHH})  with the IP-odd gauge invariant terms~\cite{Kaiser:1990yf} 
obtained from the GHLS formalism~\cite{Bando:1984pw,Bando:1987br,Bando:1987ym}.

The GHLS~\cite{Bando:1984pw,Bando:1987br,Bando:1987ym} is formulated based on the manifold $G_{\rm global} \times G_{\rm local}$ 
where $G_{\rm global}=[U(N_f)_L \times U(N_f)_R]_{\rm global}$  
and $G_{\rm local}=[U(N_f)_L \times U(N_f)_R]_{\rm local}$ associated with the GHLS.    
The dynamics of the manifold $G_{\rm global} \times G_{\rm local}$ is then described 
by the dynamical variables $\xi_L$, $\xi_R$ and $\xi_M$ forming the chiral field $U$ as 
$U=\xi_L^\dag \xi_M \xi_R$, 
together with the GHLS gauge fields $L_\mu$ and $R_\mu$. 
They transform under $G_{\rm global} \times G_{\rm local}$ 
in such a way that 
\begin{eqnarray} 
\xi_{L,R} &\to& \tilde{g}_{L,R}(x) \xi_{L,R} g_{L,R}^\dag 
\,, \\  
\xi_M &\to& \tilde{g}_L(x) \xi_M \tilde{g}_R(x)  
\,, \\  
L_\mu &\to& \tilde{g}_{L}(x) L_\mu \tilde{g}_{L}^\dag(x) + i \tilde{g}_{L}(x) \partial_\mu \tilde{g}_{L}(x) 
\, ,\\
R_\mu &\to& \tilde{g}_{R}(x) L_\mu \tilde{g}_{R}^\dag(x) + i \tilde{g}_{R}(x) \partial_\mu \tilde{g}_{R}(x) 
\,, 
\end{eqnarray}  
where $g_{L,R} \in G_{\rm global}$ and $\tilde{g}_{L,R}(x) \in G_{\rm local}$. 
The parity $(P)$ and charge conjugation $(C)$ act on $\xi_{L,R,M}$ and $L_\mu,R_\mu$ as 
\begin{eqnarray} 
&& 
\xi_L \stackrel{P}{\leftrightarrow} \xi_R 
\,, \qquad  
\xi_M \stackrel{P}{\leftrightarrow} \xi_M^\dag 
\, , \qquad 
L_\mu \stackrel{P}{\leftrightarrow} (-1)^{\mu'} R_\mu 
\,, \nonumber \\ 
&&  
\xi_L \stackrel{C}{\leftrightarrow} \xi_R^* 
\,, \qquad 
\xi_M \stackrel{C}{\leftrightarrow} \xi_M^T \, , 
\qquad 
L_\mu \stackrel{C}{\leftrightarrow} - R_\mu^T 
\,, 
\end{eqnarray}  
where $(-1)^{\mu'}$ denotes $1$ for $\mu=0$ and $-1$ for $\mu\neq 0$ regarding the corresponding Lorentz vector field.  
 The external gauge fields $\cal L$ and $\cal R$ are introduced by gauging $G_{\rm global}$ 
in the usual manner ($g_{L,R} \Rightarrow g_{L,R}(x)$).

It is convenient to introduce the following variables: 
\begin{eqnarray} 
 \omega_{L \mu} &=& \frac{1}{i} \partial_\mu \xi_{L} \xi_{L}^\dag - L_\mu + \xi_{L} {\cal L}_\mu \xi^\dag_{L}
 \,, \\ 
 \omega_{R \mu} &=& \frac{1}{i} \partial_\mu \xi_{R} \xi_{R}^\dag - R_\mu + \xi_{R} {\cal R}_\mu \xi^\dag_{R}
 \,, \\  
 \omega_{M \mu} &=& \frac{1}{i} \partial_\mu \xi_{M} \xi_{M}^\dag - L_\mu + \xi_{M} R_\mu \xi^\dag_{M}
\,, 
\end{eqnarray} 
which transform under $[G_{\rm global}]_{\rm gauged} \times G_{\rm local}$ as follows: 
\begin{eqnarray} 
 \omega_{L \mu} &\to& \tilde{g}_{L}(x) \omega_{L \mu} \tilde{g}^\dag_{L}(x)  
 \,,\\ 
  \omega_{R \mu} &\to& \tilde{g}_{R}(x) \omega_{R \mu} \tilde{g}^\dag_{R}(x)  
 \,,\\ 
 \omega_{M \mu} &\to& \tilde{g}_{L}(x) \omega_{M \mu} \tilde{g}^\dag_{L}(x) 
\,, 
\end{eqnarray} 
and under $P$- and $C$-inversions: 
\begin{eqnarray} 
&& 
\omega_{L \mu} \stackrel{P}{\leftrightarrow} (-1)^{\mu'} \omega_{R \mu}
\,, \qquad  
\omega_{M \mu} \stackrel{P}{\leftrightarrow} - (-1)^{\mu'} \xi_M^\dag \omega_{M \mu} \xi_M 
\, , \\  
&& 
\omega_{L \mu} \stackrel{C}{\leftrightarrow} - \omega_{R \mu}^T 
\,, \qquad  \qquad 
\omega_{M \mu} \stackrel{C}{\leftrightarrow} \xi_M^T \omega_{M \mu}^T \xi_M^* 
\, .
\end{eqnarray}

  The GHLS, $C$- and $P$-invariant terms  
having $\epsilon^{\mu\nu\rho\sigma}$-structure (IP-odd) 
thus take the form~\cite{Kaiser:1990yf}: 
\begin{eqnarray} 
\Gamma_{\rm GHLS}^{\rm inv}[L,R, {\cal L},{\cal R}, \xi_L^\dag\xi_M\xi_R] 
&=& \frac{N_c}{16\pi^2} \int_{M^4} \sum_{i=1}^{14} c_i {\cal L}_i 
\,, \label{sol:GHLS} 
\end{eqnarray}  
with 
\begin{eqnarray} 
{\cal L}_1 &=& 
i \, {\rm tr} [ \omega_L^3 \xi_M \omega_R \xi_M^\dag - \omega_R^3 \xi_M^\dag \omega_L \xi_M ] 
\,, \label{L1} \\ 
{\cal L}_2 &=& 
i \, {\rm tr} [ \omega_L \xi_M \omega_R \xi_M^\dag \omega_L \xi_M \omega_R \xi_M^\dag]  
\,, \\ 
{\cal L}_3 &=& 
i \, {\rm tr} [ \omega_M (\omega_L^3 + \xi_M \omega_R^3 \xi_M^\dag)] 
\,, \\ 
{\cal L}_4 &=& 
i \, {\rm tr} [ \omega_M (\omega_L \xi_M \omega_R \xi_M^\dag \omega_L + \xi_M \omega_R \xi_M^\dag \omega_L \xi_M \omega_R \xi_M^\dag) ] 
\,, \\ 
{\cal L}_5 &=& 
i \, {\rm tr} [ \omega_M (\omega_L^2 \xi_M \omega_R \xi_M^\dag + \omega_L \xi_M \omega_R^2 \xi_M^\dag + 
\xi_M \omega_R^2 \xi_M^\dag \omega_L + \xi_M \omega_R \xi_M^\dag \omega_L^2] 
\,, \\ 
{\cal L}_6 &=& 
i \, {\rm tr} [ \omega_M^2 (\omega_L \xi_M \omega_R \xi_M^\dag - \xi_M \omega_R \xi_M^\dag \omega_L) ] 
\,, \\ 
{\cal L}_7 &=& 
i \, {\rm tr} [ \omega_M (\omega_L \omega_M \omega_L - \xi_M \omega_R \xi_M^\dag \omega_M \xi_M \omega_R \xi_M^\dag) ] 
\,, \\ 
{\cal L}_8 &=& 
i \, {\rm tr} [ \omega_M^3 (\omega_L + \xi_M \omega_R \xi_M^\dag)  ] 
\,, \\ 
{\cal L}_9 &=& 
{\rm tr} [(F_L + \xi_M F_R \xi_M^\dag)(\omega_L \xi_M \omega_R \xi_M^\dag - \xi_M \omega_R \xi_M^\dag \omega_L)]
\,, \\ 
{\cal L}_{10} &=& 
{\rm tr}[(F_L + \xi_M F_R \xi_M^\dag)( (\omega_L + \xi_M \omega_R \xi_M^\dag)\omega_M -\omega_M (\omega_L + \xi_M \omega_R \xi_M^\dag) )]
\,, \\ 
{\cal L}_{11} &=& 
{\rm tr}[(F_L - \xi_M F_R \xi_M^\dag)( (\omega_L - \xi_M \omega_R \xi_M^\dag)\omega_M -\omega_M (\omega_L - \xi_M \omega_R \xi_M^\dag) )]
\,, \\ 
{\cal L}_{12} &=& 
{\rm tr} [(\xi_L {\cal F}_L \xi_L^\dag + \xi_M \xi_R {\cal F}_R \xi_R^\dag \xi_M^\dag)
(\omega_L \xi_M \omega_R \xi_M^\dag - \xi_M \omega_R \xi_M^\dag \omega_L)]
\,, \\ 
{\cal L}_{13} &=& 
{\rm tr} [(\xi_L {\cal F}_L \xi_L^\dag + \xi_M \xi_R {\cal F}_R \xi_R^\dag \xi_M^\dag)
( (\omega_L + \xi_M \omega_R \xi_M^\dag)\omega_M -\omega_M (\omega_L + \xi_M \omega_R \xi_M^\dag) )]
\,, \\ 
{\cal L}_{14} &=& 
{\rm tr} [(\xi_L {\cal F}_L \xi_L^\dag - \xi_M \xi_R {\cal F}_R \xi_R^\dag \xi_M^\dag)
( (\omega_L - \xi_M \omega_R \xi_M^\dag)\omega_M - \omega_M (\omega_L - \xi_M \omega_R \xi_M^\dag) )]
\,, \label{L14}
\end{eqnarray}
 where the notation of coefficients for ${\cal L}_{1-14}$ followed Ref.~\cite{Kaiser:1990yf}, and  
\begin{equation} 
  F_{L} = d L - i L^2 
\,, \qquad 
   F_{R} = d R - i R^2 
  \,. 
\end{equation}

We shall now introduce ``background fields" $B_L$ and $B_R$~\cite{Harvey:2007ca} 
transforming as $B_{L,R} \to g_{L,R}(x) B_{L,R} g_{L,R}^\dag(x) $ 
and relate 
them with  $\omega_L$ and $\omega_R$ as follows: 
\begin{equation} 
 B_{L} = \xi_{L}^\dag   \omega_{L}  \xi_{L} 
 \,, \qquad 
  B_{R} = \xi_{R}^\dag   \omega_{R}  \xi_{R} 
  \,. \label{replace1}
\end{equation}
 The variable $\omega_M$ and the field strengths of the GHLS fields $L$ and $R$, $F_{L,R}$,  
are then expressed in terms of the building blocks listed in Eq.(\ref{list}) 
as  
 \begin{eqnarray} 
   \omega_M &=& \xi_L ( B_L - U B_R U^\dag - i {\cal D}U U^\dag  ) \xi_L^\dag 
   \,. \label{replace2}\\ 
   F_{L(R)} &=& \xi_{L(R)} \left(  {\cal F}_{L(R)} - {\cal D}B_{L(R)} - i B_{L(R)}^2  \right) \xi_{L(R)}^\dag
\,. \label{replace3}
 \end{eqnarray}
  Putting Eqs.(\ref{replace1})-(\ref{replace3}) into Eqs.(\ref{L1})-(\ref{L14}) we rewrite 
${\cal L}_1$-${\cal L}_{14}$ in Eq.(\ref{sol:GHLS}) in terms of 
${\cal O}_1$-${\cal O}_{14}$ in Eq.(\ref{general:HHH}):   
\begin{eqnarray} 
{\cal L}_1 &=& 
{\cal O}_1 
\,, \\ 
{\cal L}_2 &=& 
{\cal O}_2 
\,, \\ 
{\cal L}_3 &=& 
{\cal O}_1 - {\cal O}_5 
\,, \\ 
{\cal L}_4  &=& 
- {\cal O}_1 + 2 {\cal O}_2 - {\cal O}_7 
\,, \\ 
{\cal L}_5 &=& 
2 {\cal O}_1 - {\cal O}_6 
\,, \\ 
{\cal L}_6 &=& 
2 {\cal O}_1 - 2 {\cal O}_2  - {\cal O}_6 + 2 {\cal O}_7  - {\cal O}_{12}
\,, \\ 
{\cal L}_7 
&=& 
 2 {\cal O}_1 - 2 {\cal O}_2 - 2 {\cal O}_5 + 2 {\cal O}_7  + {\cal O}_{13} 
\,, \\ 
{\cal L}_8 
&=& 
 2 {\cal O}_1 - 2 {\cal O}_2 - {\cal O}_5 - {\cal O}_6 + 3 {\cal O}_7 - 2 {\cal O}_{12} + {\cal O}_{13} + {\cal O}_{14}
\,, \\ 
{\cal L}_9 &=& 
- 2 {\cal O}_1 - {\cal O}_3  + {\cal O}_4 
 \,,\\ 
 {\cal L}_{10} 
 &=& 
 4 {\cal O}_1 + 2 {\cal O}_3 -2 {\cal O}_4   - 2 {\cal O}_5   - {\cal O}_6  + {\cal O}_8 + {\cal O}_9 - {\cal O}_{10}  - {\cal O}_{11}  
\,, \\ 
{\cal L}_{11} &=& 
- 2 {\cal O}_5  + {\cal O}_6  + {\cal O}_8 - {\cal O}_9 -{\cal O}_{10} + {\cal O}_{11}   
\,, \\ 
{\cal L}_{12} 
&=& 
{\cal O}_4 
\,, \\ 
{\cal L}_{13} 
&=&
 -2 {\cal O}_4 -  {\cal O}_{10} - {\cal O}_{11} 
\,, \\ 
{\cal L}_{14} 
&=& 
 - {\cal O}_{10} + {\cal O}_{11} 
\,. 
\end{eqnarray}  
 The GHLS action (\ref{sol:GHLS}) thus precisely becomes identical to the general solution in 
the HHH formulation Eq.(\ref{general:HHH})  
with the free parameters $a_1$-$a_{14}$ replaced in such a way that    
\begin{eqnarray} 
 a_1 &=&  
 c_1 + c_3 - c_4 + 2 c_5 + 2 c_6 
        + 2 c_7 + 2 c_8 - 2 c_9 + 4 c_{10} 
 \,, \\ 
 a_2 &=& 
 c_2 - c_4 - 2 c_6 - 2 c_7 - 2 c_8 
\,, \\ 
a_3 &=& 
- c_9 + 2 c_{10} 
\,, \\ 
a_4 &=& 
c_9 - 2 c_{10} + c_{12} - 2 c_{13} 
\,,\\ 
a_5 &=& 
 - c_3 - 2 c_7 -  c_8 - 2 c_{10} - 2 c_{11} 
\,, \\ 
a_6 &=& 
- 2 c_5 - c_6 -2 c_8 - c_{10} + c_{11} 
\,, \\ 
a_7 &=& 
- c_4 + 2 c_6 + 2 c_7 + 3 c_8 
\,, \\ 
a_8 &=&
c_{10} + c_{11} 
\,, \\ 
a_9 &=& 
c_{10}-c_{11} 
\,, \\ 
a_{10} &=& 
- c_{10} - c_{11} - c_{13} + c_{14} 
\,, \\ 
a_{11} &=& 
- c_{10} + c_{11} - c_{13} + c_{14} 
\,, \\ 
a_{12} 
&=& 
 - c_6 - 2 c_8 
\,, \\ 
a_{13} 
&=& 
c_7 + c_8 
\,, \\ 
a_{14} 
&=& 
c_8 
\,. 
\end{eqnarray}

%%%%%%%%%%%%%%%%%%%%%%%%%%%%%%%%%%%%%%%%%%%%%

\end{document}